\documentclass{webofc}
\usepackage[varg]{txfonts}   % Web of Conferences font
\usepackage{adjustbox}
\usepackage{subfig}
\usepackage{siunitx}
\usepackage{xcolor}

\begin{document}
\title{The use of Boosted Decision Trees for Energy Reconstruction in JUNO experiment}

\author{\firstname{Arsenii} \lastname{Gavrikov}\inst{1,2}\fnsep\thanks{\email{asgavrikov@edu.hse.ru}} \and
        \firstname{Fedor} \lastname{Ratnikov}\inst{1}\fnsep\thanks{\email{fedor.ratnikov@cern.ch}} \lastname{on behalf of the JUNO Collaboration}
        % etc.
}

\institute{HSE University, Moscow, Russia
\and
           Joint Institute for Nuclear Research, Dubna, Russia
          }

\abstract{%
	 The Jiangmen Underground Neutrino Observatory (JUNO) is a neutrino experiment with a broad physical program. The main goals of JUNO are the determination of the neutrino mass ordering and high precision investigation of neutrino oscillation properties. 
	 The precise reconstruction of the event energy is crucial for the success of the experiment. 
	  
	 JUNO is equiped with \num{17612} + \num{25600} PMT channels of two kind which provide both charge and hit time information.
	 In this work we present a fast Boosted Decision Trees model using small set of aggregated features. The model predicts event energy deposition. We describe the motivation and the details of our feature engineering and feature selection procedures.
	 We demonstrate that the proposed aggregated approach can achieve a reconstruction quality that is competitive with the quality of much more complex models like Convolution Neural Networks (ResNet, VGG and GNN).
}
\maketitle
\section{Introduction}
\label{intro}

\begin{figure}[!htb]
	\centering
	\adjincludegraphics[Clip=0 0 0 {0.404\height}, width=0.8\linewidth]{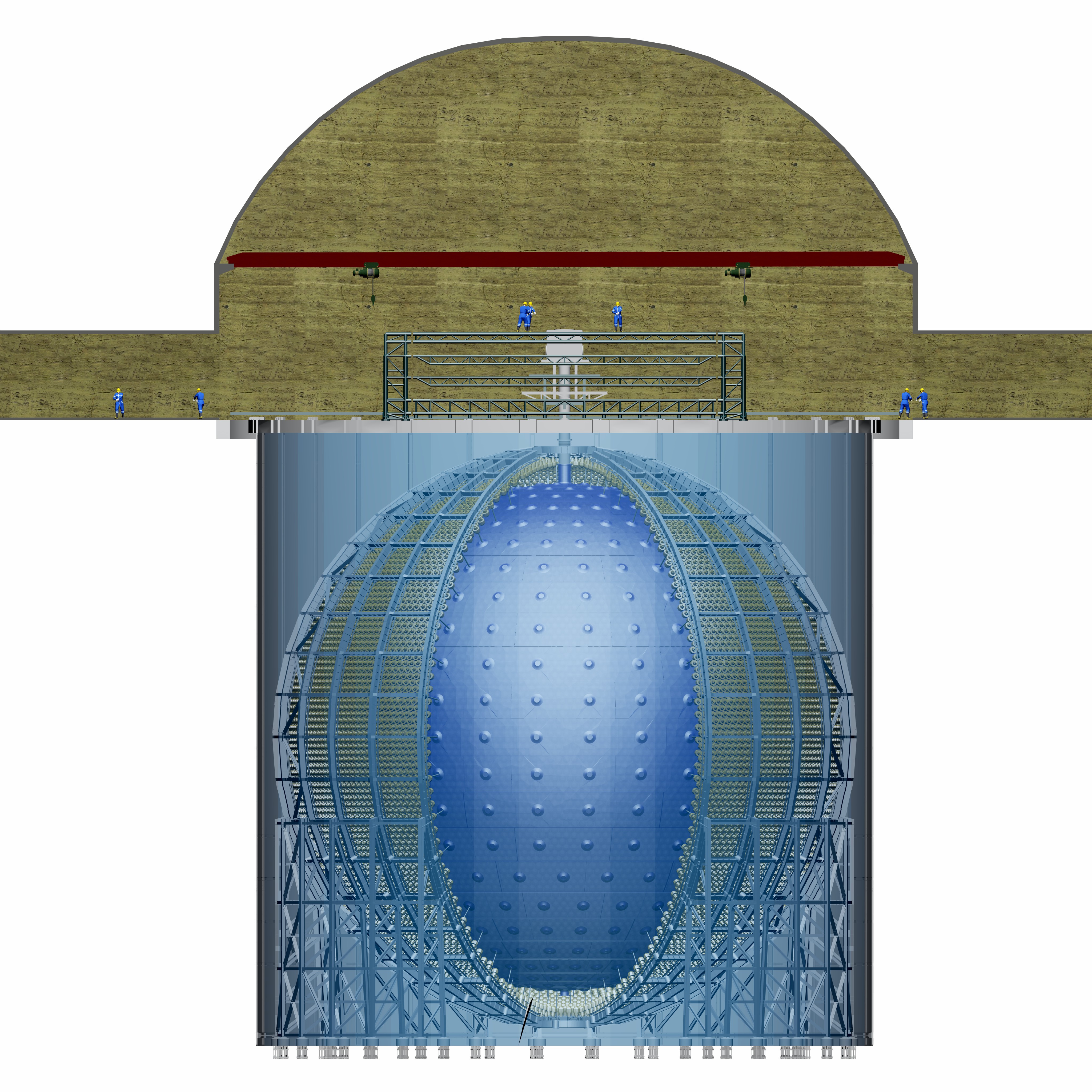}
	\caption{The central detector of JUNO. An acrylic sphere with a diameter of 35.4 meters filled with 20 kt of liquid scintillator. The detector contains \num{17612} large (20 inches) PMTs and \num{25600} small PMTs (3 inches).}
	\label{fig:detector}
\end{figure}

JUNO is a neutrino observatory under construction in southern China. Its physical program covers a wide range of problems~
\cite{JUNO}. The main goals are to determine the neutrino mass ordering and to accurately measure the parameters of neutrino oscillations $\sin^2{\theta_{12}}, \Delta m_{21}^2, \Delta m^{2}_{31}$. JUNO will detect reactor neutrinos from the Yangjiang and Taishan nuclear power plants. Simultaneously JUNO will be able to observe neutrinos from supernovae, atmospheric neutrinos, solar neutrinos and geoneutrinos.

Figure~\ref{fig:detector} shows the detector design. The detector is a transparent acrylic sphere with a diameter of 35.4 meters that is located underground in a cylindrical water pool. The sphere is filled with 20 kt of liquid scintillator. The detector is equipped with a huge number of photo-multiplier tubes (PMTs) of two types: \num{17612} large PMTs (20 inches) and \num{25600} small PMTs (3 inches). Neutrinos, which are produced in nuclear reactors, interact with the protons of the scintillator in the detector via the inverse beta-decay (IBD) channel: $\overline \nu_{e} + p \rightarrow e^{+} + n$. The scintillator then produces visible light upon the interaction of the ejected positron with the media. The amount of emitted photons is tightly related to the neutrino energy. The neutron, after some time, is captured by a hydrogen atom of liquid scintillator, producing 2.2 MeV de-excitation gammas. Thus, the time coincidence of signals from the positron and the neutron makes it possible to separate the event from backgrounds. The information collected by PMTs is used for estimation of the neutrino energy.

To resolve the neutrino mass ordering the energy resolution must be $\sigma \leqslant  3\%$ at 1 MeV, which is very close to the statistical limit corresponding to the light yield in JUNO, about 1300 detected photons (hits) at 1~MeV. The energy nonlinearity uncertainty should be < 1\%~\cite{JUNO}.

Machine Learning (ML) methods are very popular in science today, including high energy physics, in particular, neutrino experiments~\cite{Psihas} and collider experiments~\cite{Guest}. We use ML approach for energy reconstruction in the JUNO experiment. Our problem is a regression supervised learning problem. The data (time and charge information) collected by PMTs is used as input for supervised training of ML model. Earlier we demonstrated that the ML approach can have the quality required for the JUNO experiment on our data and also has the advantage of speed and ease of application~\cite{CNNs}.

In this work we use Boosted Decision Trees (BDT)~\cite{Friedman} for energy reconstruction in the energy range of 0–10 MeV covering the region of interest for IBD events from reactor electron antineutrinos. Compared to~\cite{CNNs} we designed and studied new features and achieved much better resolution with BDT, which is now comparable to the resolution of more complex models.

\section{Data description}
\label{DataDesc}

\begin{figure}[!htb]
	\centering
	\subfloat[]{\includegraphics[width=0.49\textwidth]{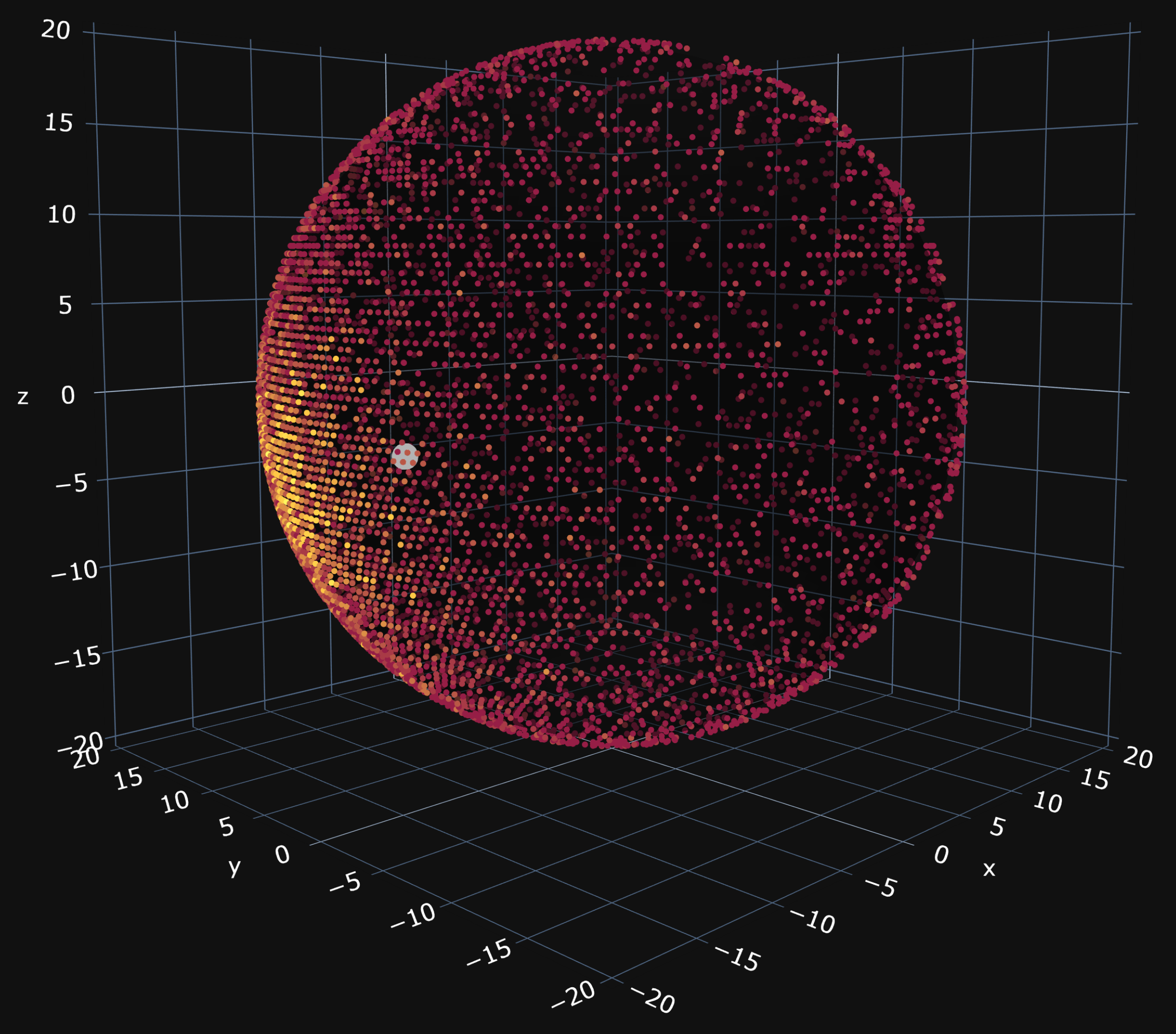}\label{fig:data_vis_ch}}
	\subfloat[]{\includegraphics[width=0.49\textwidth]{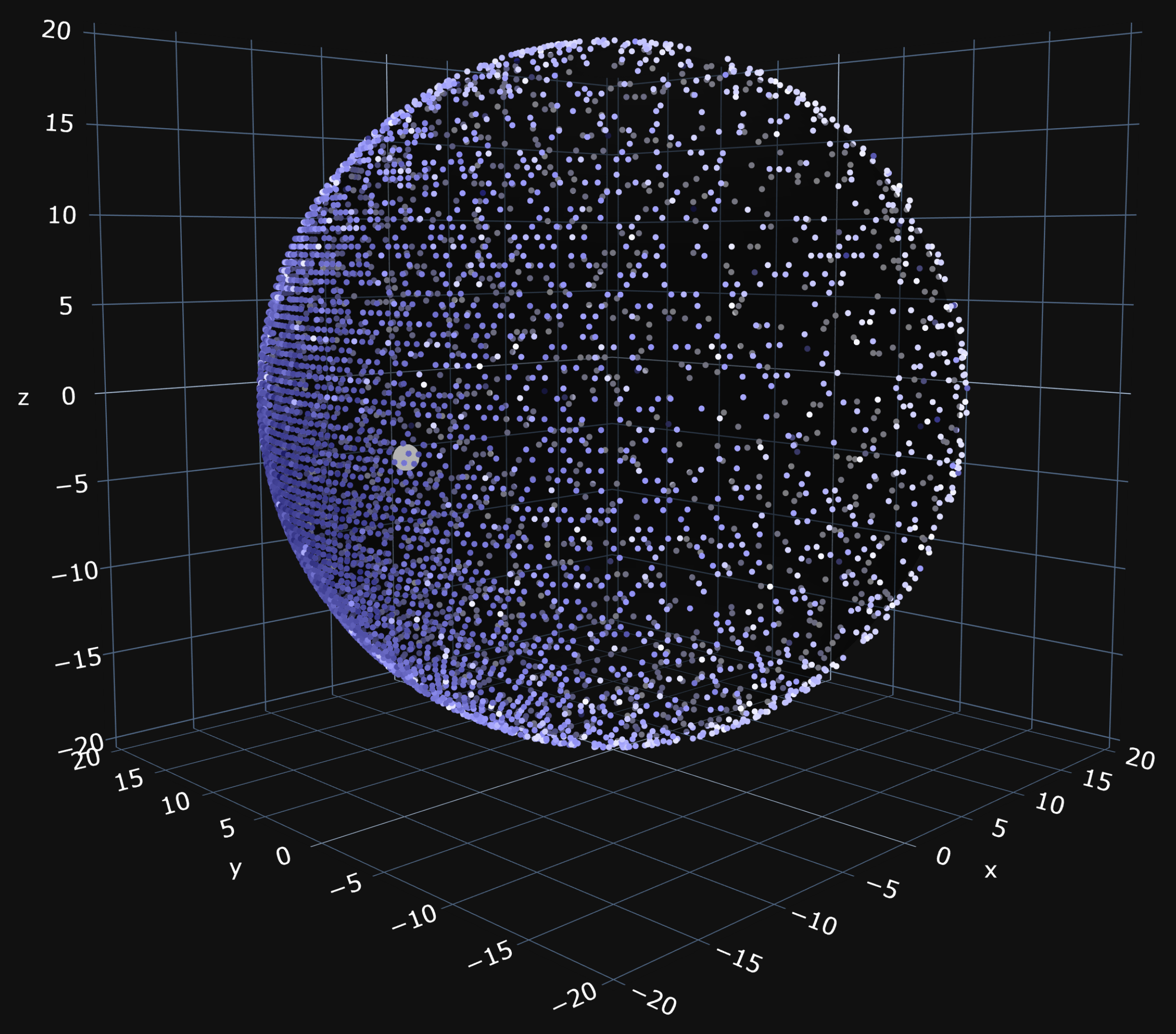}\label{fig:data_vis_ht}}
	\caption{Example of an event seen by 17612 large ($20''$) PMTs for a positron event of 5.5 MeV. Only fired PMTs are shown. In fig (a) color represents the accumulated charge in PMTs: yellow points show the channels with more hits, red points — the channels with fewer hits. In fig (b) color indicates PMT activation time --- darker blue color shows an earlier activation. The primary vertex is shown by the gray sphere.}
	\label{fig:data_vis}
\end{figure}

The dataset is generated by the full detector Monte Carlo method using the official JUNO software \cite{Huang}. The detector simulation is based on the Geant4 framework~\cite{Allison} with the geometry implemented in detail. The train and test datasets are described as follows:
\begin{enumerate}
	\item \textbf{Training dataset} consists of 5 million events, uniformly distributed in kinetic energy from 0 to 10 MeV and in the volume of the central detector (in liquid scintillator).
	
	\item \textbf{Testing dataset} consists of subsets with discrete kinetic energies of 0 MeV, 0.1 MeV, 0.2 MeV, ..., 1 MeV, 2 MeV, ..., 10 MeV. Each subset contains about 10 thousand events. This dataset is used to estimate performance after the end of training.
\end{enumerate}
Our data have four configurations: 1) without electronics effects; 
2) taking into account the transit time spread (TTS) of PMTs;
3) taking into account the dark noise (DN) of PMTs; and 4) taking into account both effects.
TTS occurs due to the stochasticity of the path of photo-electron from the photo-cathode to the anode and effectively smears the time information. DN effect gives spontaneous hits on PMTs. In further TTS and DN are always enabled if not specified otherwise.

Figure~\ref{fig:data_vis} illustrates an example of accumulated charge in the PMT channels (left) and the evolution in time of the same signal in terms of the first hit time distribution (right).

\section{Boosted Decision Trees}
\label{BDT}

BDT is the an ensemble model, where a simple and quickly learning Decision Tree (DT) model is used as the base algorithm. DTs in BDT are trained sequentially. Each subsequent DT is trained to correct errors of previous DTs in the ensemble. 
In this work we use the XGBRegressor implementation of BDT from the XGBoost library~\cite{XGB}.

DT is built recursively starting from the root node, splitting the source set into two subsets (left and right) based on the values of the input features. To build a tree, we need a principle based on which we will split the original set of objects into subsets. XGBRegressor uses Gain maximization to splitting input data into subsets.

In XGBoost the objective function contains a two parts: the training loss and the regularization term: 
\begin{equation}\label{eq:obj}
\mathcal{L}(\phi) = \sum_{i}{l \left(\hat y_{i}, y_{i} \right)} + \sum_{k} {\Omega (f_k)}
\end{equation}
A tree is penalized if the sum of the norm of values in its leaves is very large. Therefore, the regularization term is introduced here as follows:
\begin{equation}
\Omega(f) = \gamma T + \frac{1}{2}\lambda \sum^{T}_{j=0}{\omega^2_j},
\end{equation}
where $T$ is the number of leaves, $\omega_j$ are values in the leaves, $\gamma$ and $\lambda$ are numerical parameters of the regularization. In~\cite{XGB} the authors showed that the optimization of an objective function \eqref{eq:obj} reduces to maximizing of the Gain. And Gain is defined as:
\begin{equation}
{\rm Gain} = \frac{1}{2}\left[ \frac{G_l^2}{H_l^2 + \lambda} + \frac{G_r^2}{H_r^2 + \lambda} - \frac{(G_l + G_r)^2}{H^2_l + H^2_l + \lambda}\right]  - \gamma,
\end{equation}
where $G$, $H$ are the corresponding sums of the first and second derivatives of the objective function for a given partition and the indices $l$ and $r$ mean the left and right partition.

\section{Feature Engineering}
\label{FeatEng}

The basic features for the energy reconstruction are the following aggregated features: 
\begin{enumerate}
	\item Total number of detected photo-electrons (hits): \texttt{nHits}. 
	
	In the first approximation, the total number of hits is proportional to the event energy.
	\item Coordinate components of the center of charge:
	\begin{equation}
	(x_{cc},\ y_{cc},\ z_{cc}) = \mathbf{r}_{\rm cc} = \frac{\sum_i^{N_{\rm PMTs}} \mathbf{r}_{{\rm PMT}_i} n_{{\rm p.e.}, i}}{\sum_i^{N_{\rm PMTs}}n_{{\rm p.e.}, i}},
	\end{equation}
	and its radial component: 
	\begin{equation}
	R_{cc} = \sqrt{x_{cc}^2 + y_{cc}^2 + z_{cc}^2}.
	\end{equation}
	
	Coordinate components of the center of charge are rough approximations of the location of the energy deposition. These features are important for energy reconstruction since the number of hits depends on the location of the energy deposition.
	\item Coordinate components of the center of first hit time:
	\begin{equation}
	(x_{cht},\ y_{cht},\ z_{cht}) = \mathbf{r}_{\rm cht} = \frac{1}{\sum_i^{N_{\rm PMTs}} \frac{1} {t_{{\rm ht},i} + c}} \sum_i^{N_{\rm PMTs}} \frac{\mathbf{r}_{{\rm PMT}_i}} {t_{{\rm ht},i} + c},
	\end{equation}
	and its radial component: 
	\begin{equation}
	    R_{cht} = \sqrt{x_{cht}^2 + y_{cht}^2 + z_{cht}^2}.
	\end{equation}
	Here the constant $c$ is required to avoid division by zero. 
	These features bring extra information on the location of the energy deposition.
	\item Mean and standard deviation of the first hit time distributions: \texttt{ht\_mean}, \texttt{ht\_std}.
\end{enumerate}
For ML models including Boosted Decision Trees, it is often useful to engineer new features from the existing features~\cite{Heaton}. We use the following extra synthetic features:
\begin{gather}
	\gamma_{z}^{cc} = \frac{z_{cc}}{\sqrt{x_{cc}^2 + y_{cc}^2}},\
	\gamma_{y}^{cc} = \frac{y_{cc}}{\sqrt{x_{cc}^2 + z_{cc}^2}},\ 
	\gamma_{x}^{cc} = \frac{x_{cc}}{\sqrt{z_{cc}^2 + y_{cc}^2}}; \\ 
	\theta_{cc} = \arctan{\frac{\sqrt{x_{cc}^2 + y_{cc}^2}}{z_{cc}}},\ 
	\phi_{cc} = \arctan{\frac{y_{cc}}{x_{cc}}}; \\
	J_{cc} = R_{cc}^2 \cdot \sin{\theta_{cc}},\  
	\rho_{cc} = \sqrt{x_{cc}^2 + y_{cc}^2}.
\end{gather}
And some trigonometric functions of angles $\theta_{cc}$, $\phi_{cc}$: $ \sin{\theta_{cc}},\ \cos{\theta_{cc}},\ \sin{\phi_{cc}},\ \cos{\phi_{cc}}$. We also use 11 similar features for the center of first hit time.

In addition, we prepare five more features related to the location of the PMT received the maximum number of photo-electrons: \texttt{x\_max}, \texttt{y\_max}, \texttt{z\_max}, \texttt{theta\_max}, \texttt{phi\_max}, the maximum number of photons on PMT \texttt{npe\_max}, and the average number of photons on PMTs \texttt{npe\_mean}. 

Also we added the following features: \texttt{entries1}, \texttt{entries2}. Here, \texttt{entries1} is the percentage of PMTs with only 1~hit, \texttt{entries2} --- with 2~hits. And the one more feature is \texttt{nPMTs} --- the total number of fired PMTs.

\begin{figure}[!htb]
	\centering
	\subfloat[Example for a specific R.]{\includegraphics[width=0.49\textwidth]{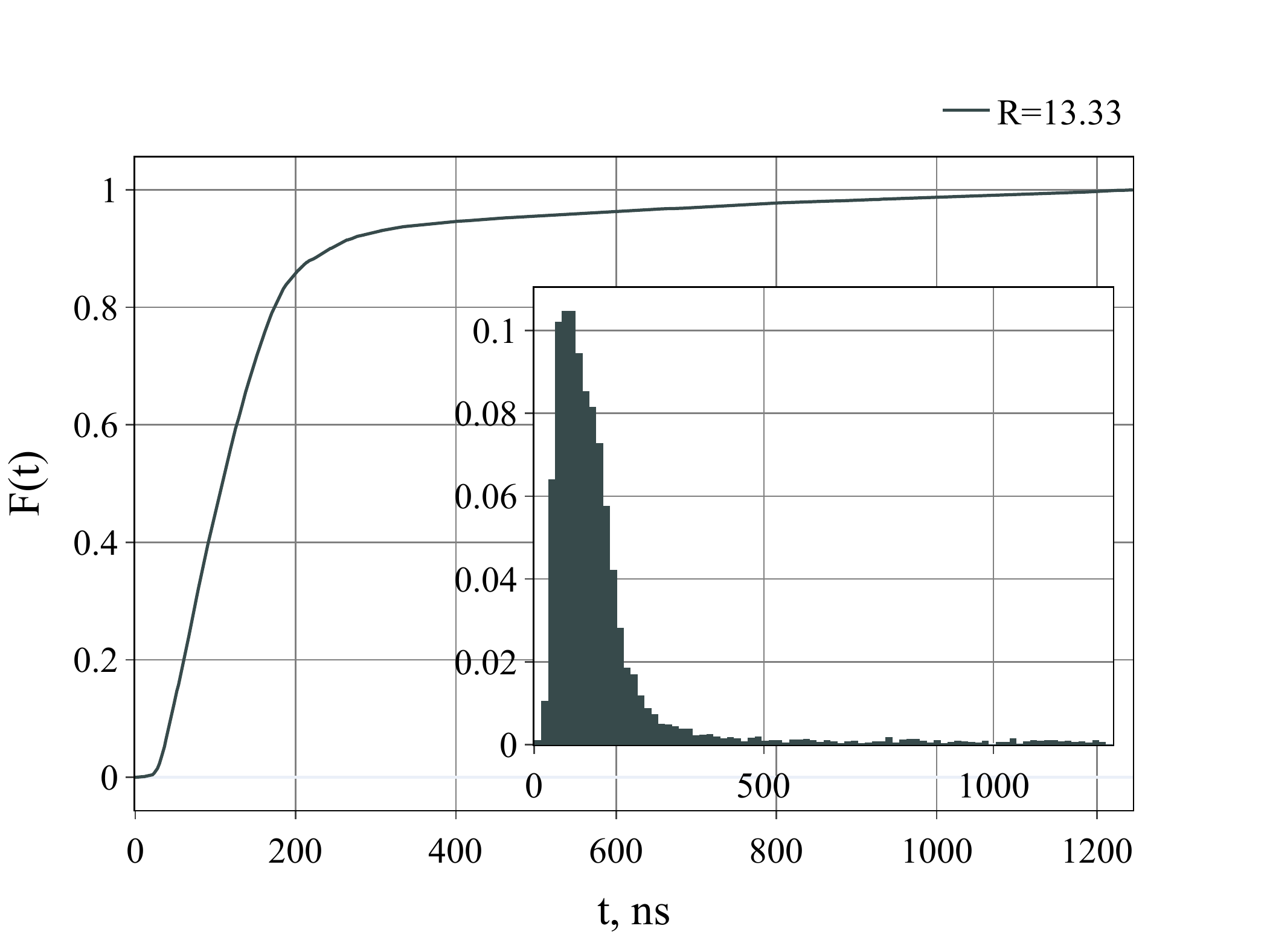}\label{fig:cdf_and_pdf_ht}}
	\subfloat[Example for different R.]{\includegraphics[width=0.49\textwidth]{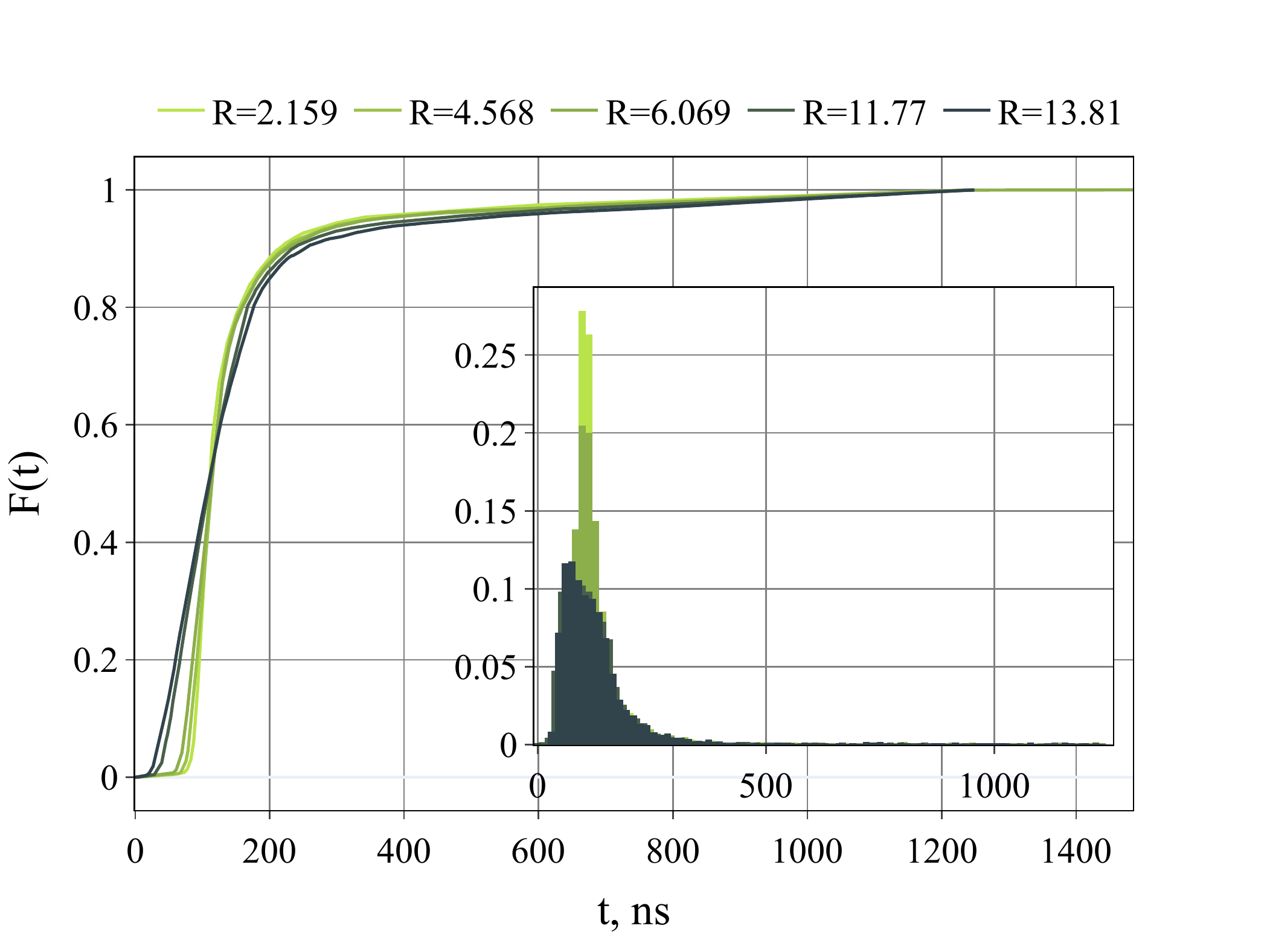}\label{fig:cdf_and_pdf_ht_diff_R}}
	\caption{Examples of CDFs and PDFs (bottom right) of the first hit time distribution for events with energy equal to 7~MeV.}
	\label{fig:ht_profiles}
\end{figure}

Now let's take a closer look at the first hit time distribution. Consider what fraction of fired PMTs received at least one photon depending on time, which is, in essence, the cumulative distribution function (CDF) of the first hit time distribution. Figure~\ref{fig:cdf_and_pdf_ht} illustrates an example of a 7~MeV event. The entire event typically lasts for about \num{1000}~ns, but the majority of photo-electrons are recorded by the PMTs in the first hundred nanoseconds and then mainly dark hits are recorded. 

In Figure~\ref{fig:cdf_and_pdf_ht} one also can see that at the beginning of events there is a short period of time $\Delta t$ during which the photons have not reached the PMTs and only dark noise is recorded.

Figure~\ref{fig:cdf_and_pdf_ht_diff_R} shows how the CDFs for the events with the energy of 7~MeV change depending on their location, closer to the edge (large R) or closer to the center of the detector (small R).
In the case where R is quite small, it takes more time for the photons to be detected by PMTs and also it takes more time to ``saturate''.

Finally, the idea is to simply decompose the entire curve into a set of percentiles, and then select those that suit best for energy reconstruction. X\%-percentile indicates how long it takes to register X\% of the first PMT hits. We use the following set of percentiles: \{1\%, 2\%, ..., 10\%, 15\%, ..., 90\%, 91\%, ..., 99\%\}.

\section{Selection of event time window}
\label{EventTimeWinSel}

One can also see in Figure~\ref{fig:ht_profiles} that the signal hits arrive within the first few hundred nanoseconds, while the dark hits form quasi-constant pedestal. Therefore, a time window can be selected, based on the data, in a way to contain mainly signal events. For this purpose we trained Boosted Decision Trees model with different bounds of window: \{75ns, 125ns, 175ns, 250ns, 500ns, 750ns, 1500ns\} and always started from $t=0$. It was trained on a 200k dataset and using all new features. 

Table~\ref{tab:bounds_results} shows the Mean Absolute Error (MAE), the Root Mean Squared Error (RMSE), Mean Absolute Percentage Error (MAPE) on the test dataset for different window bounds. The best one was found to be 500~ns, however all the windows longer than 175~ns showed similar performance. 

\begin{table}[!htb]
	\centering
	\caption{MAE, RMSE, MAPE metrics for BDT models for different window bounds.}
	\label{tab:bounds_results}
	\begin{tabular}{llll}
		\hline
		Bound, ns& MAE, MeV & RMSE, MeV & MAPE, \% \\
		\hline
		75 & 0.1019 & 0.1664 & 3.705 \\
		125 & 0.0562 & 0.0781 & 1.858 \\
		175 & 0.0505 & 0.0698 & 1.662 \\
		250 & 0.0487 & 0.0676 & 1.594 \\
		\textbf{500} & \textbf{0.0477} & \textbf{0.0660} & \textbf{1.569} \\
		750 & 0.0480 & 0.0664 & 1.585 \\
		1500 & 0.0479 & 0.0662 & 1.589 \\
		\hline
	\end{tabular}
\end{table}

\section{Feature selection}
\label{FeatSel}

Finally, we have a large set of features, but many of them are highly correlated, so we expect that a small set of them contain all the information and provide a performance close to the best possible one. Thus, the next task is to get a subset of the most informative features from the available set of all 78 features. For this purpose we use a dataset with 1M events.

\begin{figure}[!htb]
	\centering
	\includegraphics[width=\textwidth]{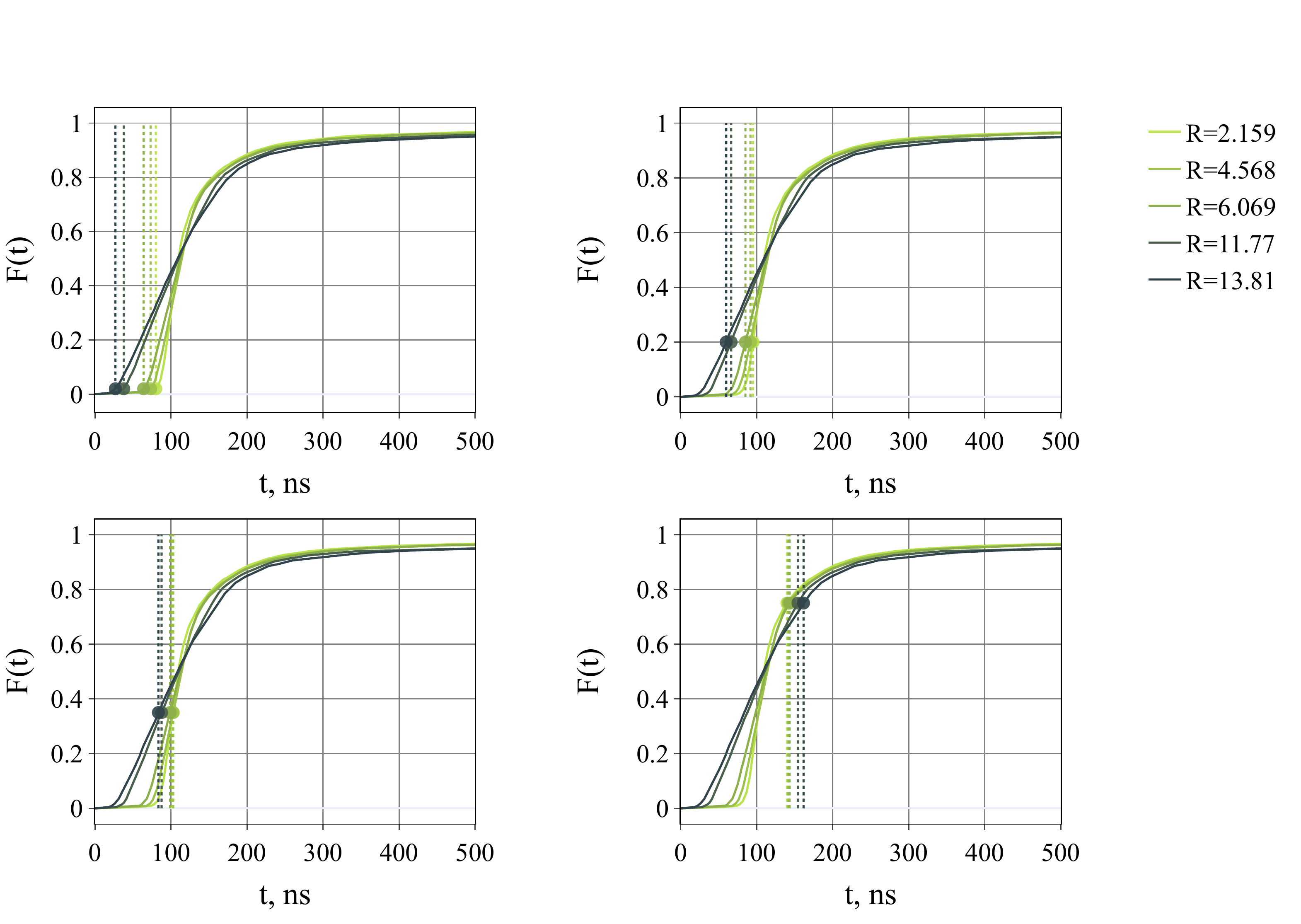}
	\caption{The selected percentiles of the CDFs of the first hit time distributions for a 7~MeV event for different radial positions: 2\% (top left), 20\% (top right), 35\% (bottom left) and 75\% (bottom right).}
	\label{fig:percentiles_vis}
\end{figure}

To select the most informative features, we use the following algorithm. First, we train a model on all features and computed RMSE on the validation dataset with 150k events. Then we take an empty list and start populating it with features. On each step we pick a feature which provide the best improvement of the model in terms of RMSE calculated on the validation dataset and put it to the end of the list. We continue while this RMSE value differs from the RMSE value for the model trained on all features by more than $\varepsilon$, chosen to be 0.0002. This procedure results to the following set of features (sorted by importance): \[\texttt{nHits},\  \texttt{ht\_20p},\ \texttt{jacob\_cc},\ \texttt{ht\_2p},\ \texttt{ht\_35p},\ \texttt{R\_cc},\ \texttt{ht\_75p}\]
Not surprisingly, for the energy reconstruction, the most informative feature is the total number of hits \texttt{nHits}, because its strongly correlated with energy, but at the same time it is hard to interpret the order of the rest features. The subset of the selected features contains \texttt{jacob\_cc} and \texttt{R\_cc}, which bring the spatial information allowing to recover the non-uniformity of detector response. 
We checked simpler combinations of features for the center of charge position. It turned out that the combination of \texttt{rho\_cc} and \texttt{R\_cc} gives the same result as \texttt{jacob\_cc} and \texttt{R\_cc}, so we have chosen them as they are more intuitive. Our final set of features is: \[\texttt{nHits},\  \texttt{ht\_20p},\ \texttt{rho\_cc},\ \texttt{ht\_2p},\ \texttt{ht\_35p},\ \texttt{R\_cc},\ \texttt{ht\_75p}\]
Figure~\ref{fig:percentiles_vis} illustrates the selected percentiles of the CDFs for an event with energy of 7~MeV and for different radial positions. As one can see \texttt{ht\_2p} contains information about the beginning of the event, that is about the moment when the number of photons PMT hits begins to grow sharply. The remaining percentiles contain information about the shape of the CDF curve and help us to separate one curve from another. The 75\% percentile is close to the moment of ``saturation'' of the CDF curve.

\section{Results}
\label{Res}

To evaluate the quality of the model, we use two metrics: resolution and bias. These metrics are obtained as a result of the Gaussian fit of $E_{\rm pred} - E_{\rm true}$ distribution. The resolution is defined as $\sigma / E_{\rm true}$ and the bias --- as $\mu / E_{\rm true}$, where $\sigma$ and $\mu$ are the standard deviation and the mean of the Gaussian distribution respectively. The performance is shown dependent on the so called visible energy, i.e.\ the maximal energy that can be converted into light: $E_{\rm vis} = E_{\rm kin} + 1.022 \text{ MeV}$. This procedure is described in more details in~\cite{CNNs}.

\begin{figure}[!htb]
	\centering
	\subfloat[Comparision of different sizes of datasets.]{\includegraphics[width=0.48\textwidth]{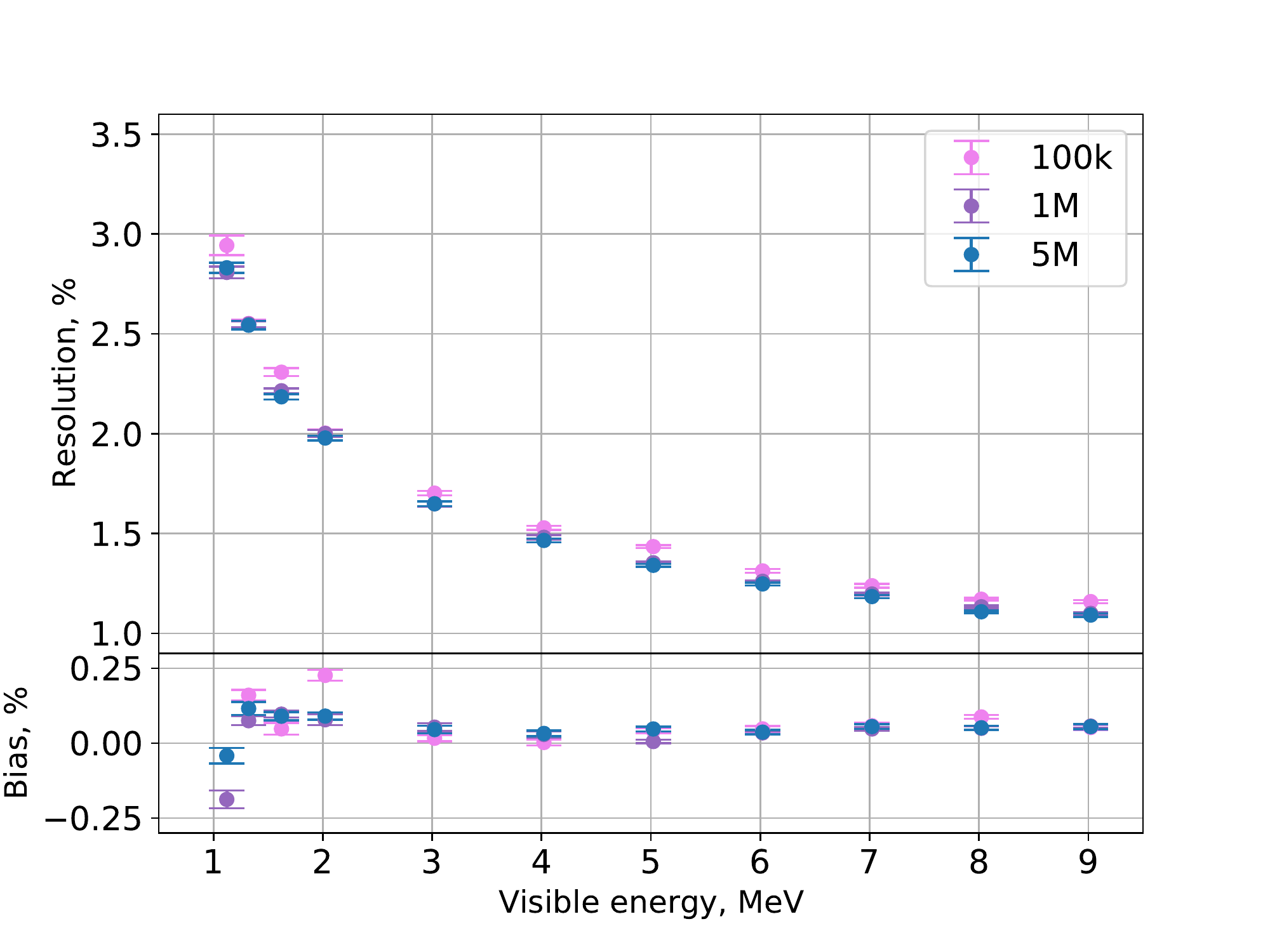}\label{fig:res_and_bias_diff_sizes_of_datasets}}
    \hspace{1em}% Space between image a and b
	\subfloat[Comparison of different TTS \& DN options for a dataset with 5M events.]{\includegraphics[width=0.48\textwidth]{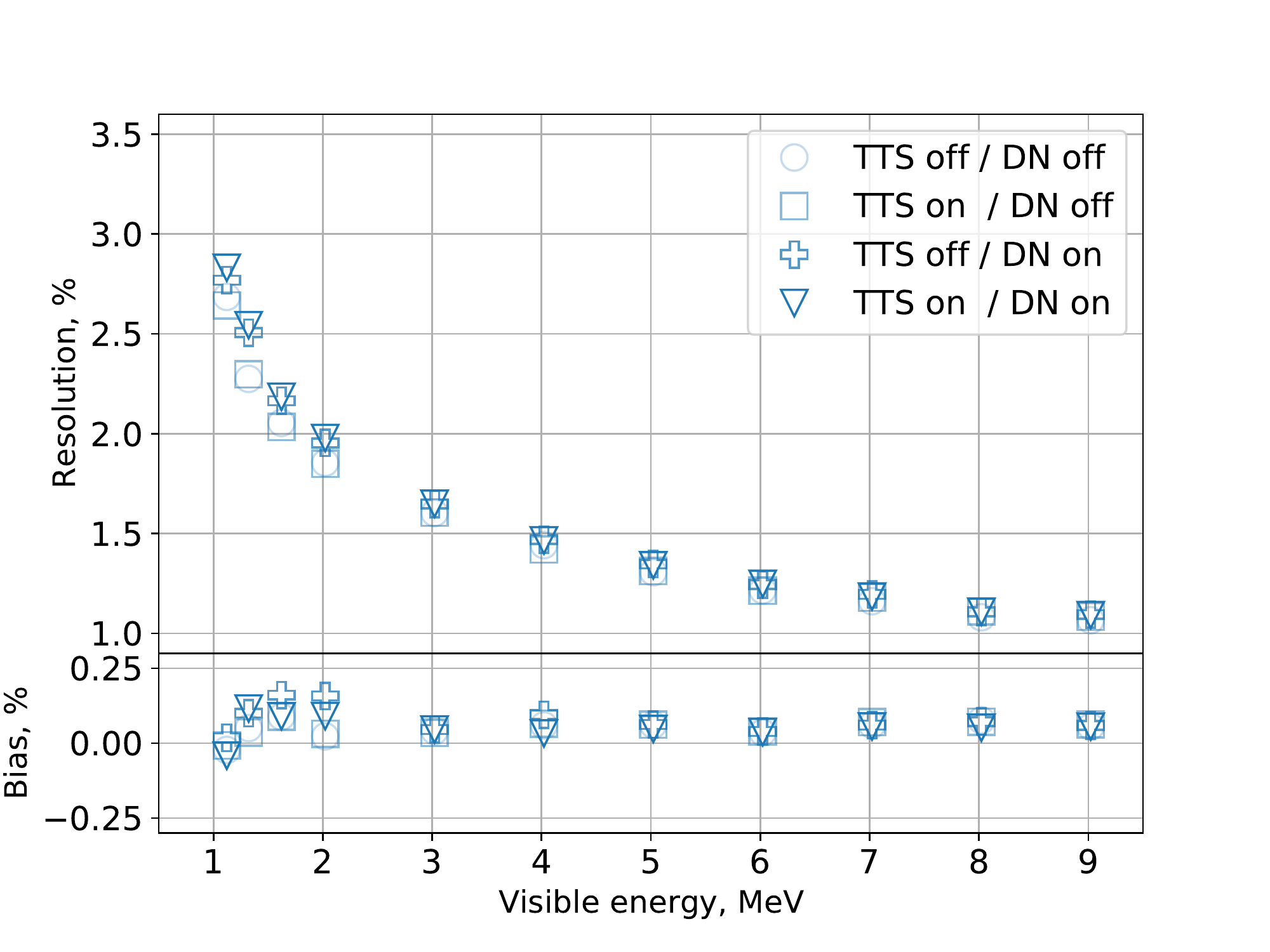}\label{fig:res_and_bias_diff_options}}
	\caption{Results of the energy reconstruction for the BDT model: resolution (upper panel) and bias (lower panel). Note that the first point corresponds to 1.122 MeV.}
	\label{fig:results_bdt}
\end{figure}

Figure~\ref{fig:res_and_bias_diff_sizes_of_datasets} illustrates the results for BDT models trained on datasets that contain different amount of events: 100k, 1M, 5M. We obtained that 1M events can provide the best possible accuracy of the model, providing only a little improvement compared to 100k events. This illustrates that fast learning is one of the advantages of the BDT model: one can get an acceptable quality already on a relatively small number of events in the dataset.

Figure~\ref{fig:res_and_bias_diff_options} shows a comparison for the BDT model trained on the 5M dataset for different options: without TTS \& DN, with TTS only, with DN only, with TTS \& DN. One can see that DN worsens the resolution, TTS --- almost does not.

\begin{figure}[!htb]
	\centering
	\includegraphics[width=0.7\textwidth]{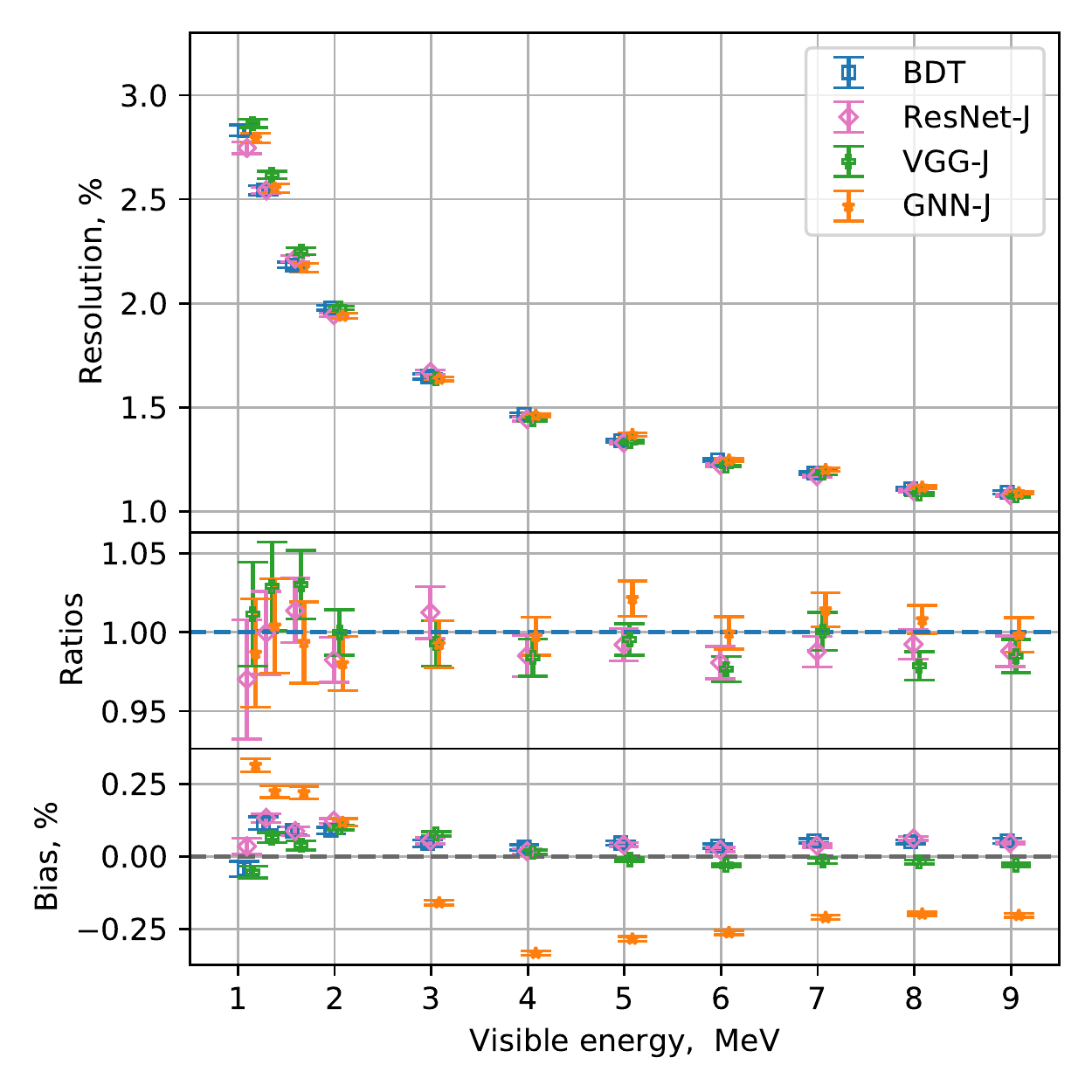}
	\caption{Energy reconstruction performance: resolution (upper panel) and bias (lower panel) obtained with BDT, ResNet-J, VGG-J and GNN-J models. The plots are offset along X-axis within $\pm$0.06 MeV for better readability. Note that the first point corresponds to 1.122 MeV.}
	\label{fig:res_and_bias_all_models}
\end{figure}

A comparison of the BDT model with other more complex deep learning models (ResNet, VGG, and GNN)~\cite{CNNs} is shown in Figure~\ref{fig:res_and_bias_all_models}. All the models are trained on the dataset with 5M events. We can see that the performance of the energy reconstruction with BDT model is practically similar to the complex deep learning models. At the same time the computations required for training and prediction are much faster due to the minimalistic nature of BDT.

\section{Summary}
\label{Sum}

In this work we have presented the use of Boosted Decision Trees for energy reconstruction in the JUNO experiment in the relevant energy range. We have designed and investigated a large set of features and have selected a small subset providing the performance nearly equal the one obtained with the full set of features. Using such a minimalistic and fast model as BDT we achieved a performance similar to the one of more complex models like ResNet, VGG, GNN.

\section*{Acknowledgements}
\begin{acknowledgement}
We are immensely grateful to Yury Malyshkin for his invaluable contribution to this work. We would like to thank Weidong Li, Jiaheng Zou, Tao Lin, Ziyan Deng, Guofu Cao and Miao Yu for their tremendous contribution to the development of JUNO offline software and to Xiaomei Zhang and Jo\~ao Pedro Athayde Marcondes de Andr\'e for production of the MC samples. We are grateful to N.~Kutovskiy, N.~Balashov for providing an extensive IT support and computing resources of JINR cloud services~\cite{Baranov}.
Fedor Ratnikov is supported by the Russian Science Foundation under grant agreement \textnumero{17-72-20127}.
\end{acknowledgement}

\end{document}